\documentclass[twocolumn,showpacs,preprintnumbers,amsmath,amssymb]{revtex4}

\usepackage{graphicx}
\usepackage{dcolumn}
\usepackage{bm}

\begin{document}

\preprint{}

\title{Observation of the Stark-Tuned F$\ddot{\mathrm{o}}$rster Resonance between Two Rydberg atoms}
\author{I.~I.~Ryabtsev}
  \email{ryabtsev@isp.nsc.ru}
\author{D.~B.~Tretyakov}
\author{I.~I.~Beterov}
\author{V.~M.~Entin}
\affiliation{Institute of Semiconductor Physics, Prospekt Lavrentyeva 13, 630090 Novosibirsk, Russia }

\date{18 February 2010}

\begin{abstract}
Cold atoms in highly excited Rydberg states are promising candidates to implement quantum logic gates of a quantum 
computer via long-range dipole-dipole interaction. Two-qubit gates require a controlled interaction of only two close 
Rydberg atoms. We report on the first spectroscopic observation of the resonant dipole-dipole interaction between two 
cold rubidium Rydberg atoms confined in a small laser excitation volume. The interaction strength was controlled by fine 
tuning of the Rydberg levels into a F$\ddot{\mathrm{o}}$rster resonance using the Stark effect. The observed 
resonance line shapes are in good agreement with numerical Monte-Carlo simulations.
\end{abstract}

 \maketitle

Quantum computers are among of the most intriguing challenges for researchers in different areas of physics. As no fully working quantum computer 
has been reported yet, many approaches are being developed simultaneously [1,2]. One promising approach is 
a quantum computer based on neutral alkali-metal atoms trapped in optical lattices or dipole traps [2]. Each atom represents 
a qubit whose quantum states are the two hyperfine sublevels of the ground state. 

The most challenging task with neutral atoms is to implement the two-qubit quantum logic gates needed to entangle 
the qubits. Such gates require a controlled interaction between qubits separated by several microns. As ground-state atoms interact 
very weakly at such distances, it has been proposed [3,4] to excite atoms to high Rydberg states with principal quantum 
number $n\gg 1$. This provides strong dipole-dipole interaction (DDI) between qubits ($\sim n^4$) [5]. 
 
Two basic proposals consider a short-term DDI of two close Rydberg atoms [3] or laser excitation of only one Rydberg atom 
in a mesoscopic ensemble (dipole blockade) [4]. A combination of these ideas has resulted recently in the first 
implementation of the controlled-NOT quantum gate with fidelity 0.73 [6] and of the entanglement with fidelity 0.75 [7] 
using the dipole blockade at laser excitation of two Rydberg atoms separated by 10 or 4~$\mu$m. Both experiments actually 
employ quasiresonant dipole-dipole interaction at quasi-F$\ddot{\mathrm{o}}$rster resonance, instead of the Stark-tuned 
F$\ddot{\mathrm{o}}$rster resonance originally proposed in Ref. [4]. 

F$\ddot{\mathrm{o}}$rster resonant energy transfer appears due to DDI of the atoms
excited to a level that lies midway between two other levels [8]. For some levels it can be precisely
tuned with an electric field via the Stark effect [9]. Stark-tuned F$\ddot{\mathrm{o}}$rster resonance 
is more flexible in controlling the interaction strength, otherwise one 
needs to change the Rydberg states or interatomic distance. Stark-tuned F$\ddot{\mathrm{o}}$rster resonance is also 
advantageous as it allows for direct measurement of the interaction strength between two atoms by recording its 
spectrum in the electric-field scale and subtracting the nonresonant background signal [10].
However, such a resonance was never observed for two Rydberg atoms, to the best of our knowledge.

In this Letter we report on the first experimental observation of the Stark-tuned F$\ddot{\mathrm{o}}$rster resonance 
between a definite number (2$-$5) of closely spaced Rydberg atoms.
Such a study is a prerequisite for future quantum gates that employ 
F$\ddot{\mathrm{o}}$rster resonances controlled with a weak dc electric field. In particular, line-shape analysis is important, 
as the resonance can be broadened by various sources of decoherence and thus may affect fidelity of quantum gates. 
Coherent or incoherent evolution of the interacting Rydberg atoms, while experiencing various decoherence processes, should be precisely investigated.  
Electrical control of the F$\ddot{\mathrm{o}}$rster resonance can be used to adjust the phase of the collective wave function 
in quantum phase gates or to implement the dipole blockade [11].  

Experiments with few Rydberg atoms require special techniques for their manipulation and detection. 
We have developed a method to separately measure the signals from $N=1-5$ of the detected Rydberg atoms [10]. 
Our technique uses a selective field ionization (SFI) detector [5] with a channel electron multiplier and relies on post-selecting the signals.  
In this work we applied it to cold Rb Rydberg atoms in a magneto-optical trap (MOT).

The F$\ddot{\mathrm{o}}$rster resonance under study is the resonant energy transfer 
$\mathrm{Rb}\left( {37P_{3/2}} \right) + \mathrm{Rb}\left( {37P_{3/2}}  
\right) \to \mathrm{Rb}\left( {37S_{1/2}}  \right) + \mathrm{Rb}\left( {38S_{1/2}}  \right)$ 
due to DDI of two or more Rb Rydberg 
atoms in a small laser excitation volume. The initial energy detuning $\Delta = \left[ {2E\left( {37P_{3/2}}  
\right) - E\left( {37S_{1/2}}  \right) - E\left( {38S_{1/2}}  \right)} 
\right]/h$ in a zero electric field is 103 MHz; $\Delta$ becomes zero at 1.79 V/cm.

The experiments were performed with cold $^{85}$Rb atoms in a MOT of standard 
configuration. Typically 10$^{5}$-10$^{6}$ atoms were trapped 
in a 0.5$-$0.6~mm diameter cloud. 
The electric field for SFI was formed by two stainless-steel plates 
with holes and meshes for passing the vertical laser cooling beams
and the electrons to be detected. Channel electron multiplier output pulses from the 37\textit{S} and 
(37\textit{P}+38\textit{S}) states were detected with two independent gates 
and sorted according to the number of detected Rydberg atoms \textit{N}.

The excitation of Rb atoms to the 37\textit{P} Rydberg state was 
realized in four steps: (i) $5S \to 5P_{3/2} $ with a 780~nm cw cooling 
laser; (ii) $5P_{3/2} \to 8S$ with a 615~nm pulsed Rh 6G dye laser; 
(iii) $8S \to \left( {6P,7P} \right) \to 6S$ spontaneous cascade decay 
during 200 ns; and (iv) $6S \to 37P$ with a 743~nm pulsed Ti:sapphire 
laser. The pulsed lasers had a pulse width of 50~ns at a repetition rate of 
5~kHz.

A small Rydberg excitation volume was formed using a crossed-beam geometry [12]. 
The two pulsed laser beams were focused to a waist of 9$-$10~$\mu $m in diameter 
and intersected at right angles inside the cold atom 
cloud. The laser intensities were adjusted to obtain 
about one Rydberg atom excited per laser pulse on average. Numerical 
simulation for these intensities gave the effective volume of $ \simeq
$5800~$\mu$m$^{3}$ (a cubic volume of 18~$\mu$m in size). 

Microwave spectroscopy was applied for diagnostics of the 
electromagnetic fields in the excitation volume: dc electric field was calibrated with 
0.2$\%$ uncertainty using Stark spectroscopy of the microwave transition 
37\textit{P}$_{3/2 } \to 37\textit{S}_{1/2}$; MOT magnetic field was not switched off, 
but the microwave probing allowed us to align the excitation point to a nearly zero magnetic field [12].

We observed formation of $2-5$ ions due to photoionization at laser excitation. 
The ions were removed by a 5~V/cm electric-field pulse of 2~$\mu$s duration. 
After this pulse, the electric field decreased to a variable value of 1.7$-$2.1~V/cm and acted for 3~$\mu$s
until the SFI pulse.

The spectra of the F$\ddot{\mathrm{o}}$rster resonances were separately recorded for 
various \textit{N}=1$-$5. The measured signals are the fraction of atoms in the final 37\textit{S} state

\begin{equation}
\label{eq1}
S_{N} = \frac{{n_{N} \left( {37S} \right)}}{{n_{N} \left( {37P} \right) + 
n_{N} \left( {37S} \right) + n_{N} \left( {38S} \right)}},
\end{equation}

\noindent
where $n_{N} \left( {nL} \right)$ is the total number of \textit{nL} Rydberg 
atoms detected by SFI during the accumulation time for the particular case 
of \textit{N} detected Rydberg atoms. Upon complete coherent transfer from the 37\textit{P}
state to the 37\textit{S} and 38\textit{S} states, $S_N$ reaches a maximum value of 0.5,
as both final states are populated with equal probability.
An incoherent transfer to those states leads to $S_N \leq$0.25.

The  F$\ddot{\mathrm{o}}$rster resonance spectra recorded for \textit{N}=1$-$5 after a free interaction time 
$t_{0}$=3~$\mu $s are shown in Fig.~1(a). Each point was averaged over $\simeq$10$^5$ laser pulses.
The laser polarization was oriented along the dc electric 
field to provide excitation of only the 37\textit{P}$_{3/2}$($\vert M_{J}\vert$=1/2) atoms 
from the intermediate 6\textit{S} state. In this case a single resonance at 1.79~V/cm
is observed. The height and width of the resonance grow with \textit{N}, as expected from simple theoretical considerations [10].
However, precise comparison between theory and experiment should be based on
an adequate theoretical model for few Rydberg atoms. In what follows, we 
calculate the theoretical spectra and then compare them with the experiment.

\begin{figure}
\includegraphics[scale=0.6]{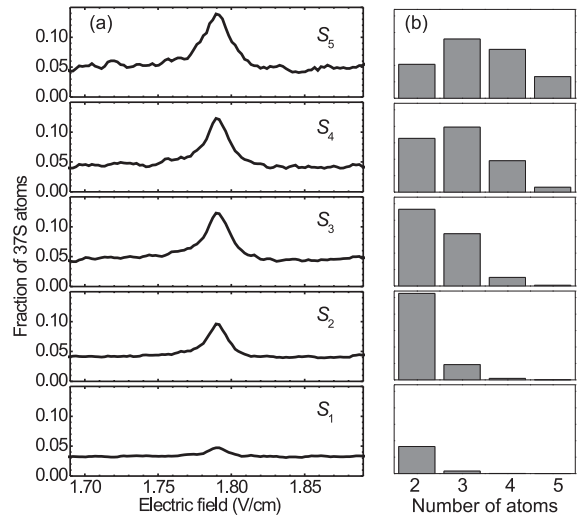}
\caption{\label{Fig1}(a) Experimental spectra $S_{1}-S_5$ of the F$\ddot{\mathrm{o}}$rster resonance
$\mathrm{Rb}\left( {37P_{3/2}} \right) + \mathrm{Rb}\left( {37P_{3/2}}  \right) \to \mathrm{Rb}\left( {37S_{1/2}}  \right) + 
\mathrm{Rb}\left( {38S_{1/2}}  \right)$
 for 1$-$5 detected Rydberg atoms. (b) Theoretical probability distributions given 
by Eqs.~(2) and (3) for the number of actually interacting Rydberg atoms.}
\end{figure}

The first item to be discussed is, When we observe the spectrum $S_N$ for \textit{N} detected Rydberg atoms,
does it really correspond to the spectrum $\rho _{N} $ for the interaction of exactly \textit{N} atoms? 
In our previous paper [10] we have shown that for an ideal SFI detector the signal $S_N$ indeed gives 
the true spectrum $\rho _{N} $. For the nonideal detector, which detects fewer atoms than have actually interacted, 
various $\rho _{i} $ contribute to $S_{N} $ to a degree that depends on the mean number of the detected Rydberg atoms. The spectra \textit{S}$_{N}$ 
are thus a mixture of the spectra from the larger numbers of actually interacting atoms $i \ge N$ [10]:

\begin{equation}
\label{eq2}
S_{N} = \,\rho  + e^{ - \bar {n}\left( {1 - T} \right)}\sum\limits_{i = 
N}^{\infty}  {\rho _{i} \frac{{\left[ {\bar {n}\left( {1 - T} \right)} 
\right]^{i - N}}}{{\left( {i - N} \right)!}}} .
\end{equation}

\noindent
Here $\rho  $ is a nonresonant background signal due to blackbody-radiation-induced transitions and background collisions, 
$\bar {n}$ is the mean number of Rydberg atoms excited per laser pulse, and \textit{T} is the detection efficiency of the SFI detector. 
The mean number of Rydberg atoms detected per laser pulse is $\bar {n}T$. In our experiment it was measured to be $\bar{n}T=0.65\pm 0.05$. 
According to the method developed in Ref. [10], this value along with the measured relationship between the one- and two-atom resonance amplitudes 
$\alpha = \left( {S_{1} - \rho } \right)/\left( {S_{2} - \rho }  \right) =0.27\pm 0.03$ 
 gives the unknown values of $\bar {n} =\left[ \alpha /\left( {1 - \alpha}  \right) + \bar 
{n}T \right] =1.05\pm 0.04$ and $T=\left( {65 \pm 5} \right)\% $. 

Equation (2) cannot be directly used to describe $S_N$ in Fig.~1(a) as the fourth excitation step uses a broadband laser radiation.
The dipole blockade effect is avoided with such radiation, however unwanted excitation to both fine sublevels of the $37P$ state 
is produced. Atoms in the $37P_{1/2} $ state do not interact
but contribute to the number of the detected atoms. This leads to an 
additional mixing of the multiatom spectra, which can be taken into account 
by replacing $\rho _{i} $ in Eq.(\ref{eq2}) with a convolution of the probability to 
excite $i \ge 2$ atoms in a laser shot and then to find $k \ge 2$ of these 
in the interacting $37P_{3/2} $ state ($k \ge 2$ as we need at least two atoms to interact):

\begin{equation}
\label{eq3}
\rho _{i} \to \left[ {\sum\limits_{k = 2}^{i} {\rho _{k} \left( {p_{3/2}}  
\right)^{k}\left( {p_{1/2}}  \right)^{i - k}\frac{{i\,!}}{{k\,!\left( {i - 
k} \right)!}}}}  \right].
\end{equation}

\noindent
Here $p_{3/2} $ and $p_{1/2} $ are the relative probabilities to excite the 
$37P_{3/2} $ and $37P_{1/2} $ atoms. Calculations have shown 
that at our laser intensities the $6S \to 37P$ transition strongly saturates and these atoms are excited with 
almost equal probabilities of $p_{3/2}\approx 0.52$ and $p_{1/2}\approx 0.48$.

The resulting theoretical distributions over $\rho _{i} $, calculated with Eqs.~(2) and (3) 
for various $S_{N} $, are presented as histograms in Fig.~1(b). 
These histograms show that the  $S_{1} $ and $S_{2} $ spectra almost completely originate from the 
interaction of two Rydberg atoms. To the best of our knowledge, this is the first observation of the 
Stark-tuned F$\ddot{\mathrm{o}}$rster resonance for two Rydberg atoms, which is the central result 
of this Letter. The method used thus paves the way to a detailed investigation of two-atom interactions, 
even when the SFI detection efficiency is far below 1. In particular, the line-shape analysis  
is necessary to reveal the effective coherence time at the F$\ddot{\mathrm{o}}$rster resonance
to characterize the future quantum gates.

For the line-shape analysis we need  to calculate various  $\rho _{i} $. 
We applied a simplified theoretical model that considered only interactions of Rydberg atoms
in the identical 37\textit{P}$_{3/2}$($\vert M_{J}\vert$=1/2) state.
The operator of the DDI between two such atoms \textit{a} and \textit{b} is then reduced to

\begin{equation}
\label{eq4}
\hat {V}_{ab} = \frac{{\hat {d}_{a} \hat {d}_{b}} }{{4\pi \varepsilon _{0} 
}}\left( {\frac{{1}}{{R_{ab}^{3}} } - \frac{{3\,\,Z_{ab}^{2}} }{{R_{ab}^{5} 
}}} \right),
\end{equation}

\noindent
where $\hat {d}_{a,b} $ are the \textit{z} components of the dipole-moment 
operators of the two atoms, $R_{ab} $ is the distance between the atoms, 
$Z_{ab} $ is the \textit{z} component of the vector connecting the two atoms 
(the \textit{z} axis is chosen along the dc electric field), and $\varepsilon 
_{0}$ is the dielectric constant. We have performed numerical Monte-Carlo 
simulations for \textit{i}=2$-$5 interacting Rydberg atoms, randomly positioned in a small
cubic volume. In this approach, the time evolution 
of all possible quasimolecular states is obtained by numerically solving the 
Schr$\ddot{\mathrm{o}}$dinger equation. We accounted for all possible 
binary resonant interactions between \textit{i} atoms, as well as the 
always-resonant exchange interactions that may broaden the resonance. The 
initial positions of \textit{i} atoms were averaged over 500 random 
realizations. A similar approach was used in Ref. [13].

\begin{figure}
\includegraphics[scale=0.52]{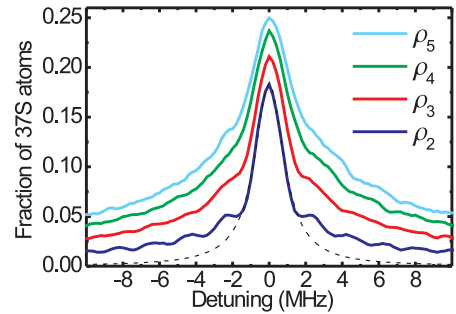}
\caption{\label{Fig2}(color online). Theoretical spectra $\rho_i$ of the F$\ddot{\mathrm{o}}$rster resonance
$\mathrm{Rb}\left( {37P_{3/2}} \right) + \mathrm{Rb}\left( {37P_{3/2}}  \right) \to \mathrm{Rb}\left( {37S_{1/2}}  \right) + 
\mathrm{Rb}\left( {38S_{1/2}}  \right)$ 
obtained by numerical Monte-Carlo simulations for $i=2-5$ atoms interacting for 0.515~$\mu$s and randomly positioned in 
a $18\times18\times18$~$\mu$m$^3$ cubic volume. The dashed line is an attempt of the Lorentz fit for $\rho_2$.}
\end{figure}

Figure~2 shows the results of our calculations for 2$-$5 atoms interacting 
for $t_{0}$=0.515~$\mu$s and randomly positioned in a $18 \times 18 
\times 18$~$\mu$m$^{3}$ cubic volume. The interaction time was chosen to fit the widths of the
resonances in Fig.~1(a), while the volume corresponds to the conditions of our experiment. The respective 
Rydberg atoms density is low and varies in the $(3-9)\cdot 10^8$~cm$^{-3}$ range.
The short interaction time allowed us to simplify the 
calculations by ignoring the effective lifetimes (30$-$40~$\mu$s 
[14]), the motion of the atoms, and the hyperfine structure (for $^{85}$Rb the 
estimated splittings are 380, 350, and 60~kHz for 37\textit{S}, 
38\textit{S}, and 37\textit{P}$_{3/2}$ states, respectively).  
Accounting for the hyperfine structure is impossible due to the enormously large number 
of quasimolecular states, which in addition are strongly mixed by the electric field.

The theoretical spectra in Fig.~2 highlight several features. (i)~At short 
interaction times the ultimate full width at half maximum (FWHM) is defined 
by the inverse interaction time $1/(t_{0})\approx$1.94~MHz 
for \textit{i}=2. It is mainly a Fourier-transform limited width, which is larger 
than the estimated 0.3~MHz energy of DDI at an average distance of 10~$\mu $m. 
The probability for two atoms to interact at much shorter distances is low,
and the resonance is narrow in spite of the spatial averaging.
(ii)~For $i > 2$ the resonances broaden due to increase in 
the average energy of DDI. Averaging over the atom positions forms a 
resonance with broad wings and a cusp on the top, which is similar to that 
observed in atomic beams [10,15]. (iii)~The resonance amplitude $\rho _{i} 
\left( {\Delta = 0} \right)$ tends to saturate at the 0.25 value. This is 
due to the loss of coherence and washing out of the Rabi-like population 
oscillations upon spatial averaging. (iv) A dashed line in Fig.~2 is 
an attempt to fit the $\rho _{2} $ spectrum with a Lorentz profile. It is seen that when we match 
the central part, the resonance wings are much broader than the Lorentz 
ones, due to rare interactions at very short distances. The commonly used Lorentz fit is thus inadequate for 
precise comparison between theory and experiment, especially when a constant 
background signal is present in the experimental spectra. 

\begin{figure}
\includegraphics[scale=0.6]{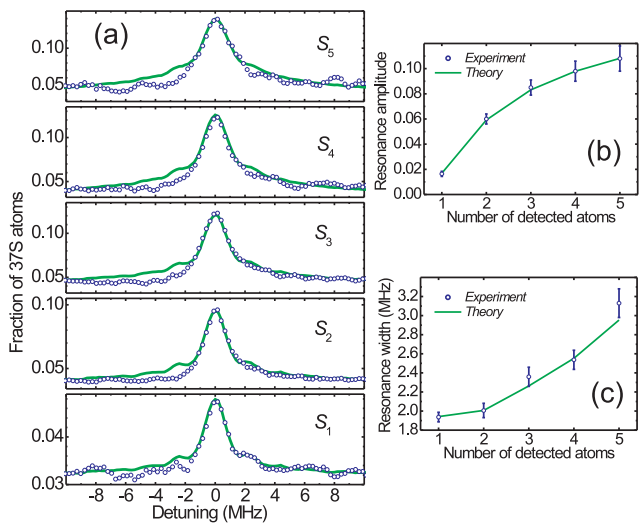}
\caption{\label{Fig3} (color online). (a) Comparison between theoretical (solid lines) and experimental 
(open circles) spectra of the F$\ddot{\mathrm{o}}$rster resonance. Theory uses the spectra of Fig.~2 and takes into account 
the mixing of these spectra due to finite detection efficiency and excitation of the noninteracting $37P_{1/2} $  atoms. 
Experiment corresponds to Fig.~1(a). (b) Theory and experiment for the resonance amplitude. 
(c) Theory and experiment for the resonance width. }
\end{figure}

Now we are able to analyze the line shapes in Fig.~1(a). In the $S_{1} $ spectrum, the FWHM in 
the electric-field scale is $16.4\pm 0.3$~mV/cm, which corresponds to $1.94\pm 0.04$~MHz. 
As the $S_{1} $ spectrum does not saturate, its width is 
defined by the effective interaction time that should be 0.515~$\mu $s 
according to theory. This time is shorter than $t_{0}$=3~$\mu$s we set up in the experiment. 
The reasons for this discrepancy can be understood as follows. The free interaction time 
contributes about 0.3~MHz. The overall hyperfine structure contributes 
another 0.8~MHz. The inhomogeneous MOT magnetic field gives about 0.2~MHz. 
The remaining 0.7~MHz can be attributed to the parasitic ac electric fields 
of 5$-$6~mV/cm due to stray fields and ground loops. This analysis also agrees with the width of 
the microwave spectrum observed in the 1.79~V/cm electric field.

Our Monte-Carlo simulations have shown that all above broadenings can be accounted for by simply 
reducing the interaction time to 0.515~$\mu $s instead of 3~$\mu $s. A possible explanation is that any small level splitting, 
which is unresolved in the observed spectrum, adds to the total resonance width. As the resonance 
width is given by the inverse effective interaction time, this time should decrease when the splitting is taken into account. 
We conclude that the unresolved hyperfine, Zeeman, and Stark structures of the 
F$\ddot{\mathrm{o}}$rster resonance lead to a decrease of the effective interaction time and are thus the main sources of decoherence.
Fast quantum gates should be implemented at much shorter interaction times.

The final theoretical spectra are shown as solid lines in 
Fig.~3(a). The open circles are the experimental data of Fig.~1(a) in the detuning scale. The nonresonant 
background level was added to the theory in order to fit the experiment at the far wings of the resonance and to correctly determine the resonance height.
Figure~3(a) demonstrates the good agreement between the theoretical and 
experimental line shapes for all \textit{N}. The wings of the experimental  two-atom spectra $S_{1} $ and $S_{2} $ even 
reproduce some of the coherent features due to population oscillations. The slight asymmetry in the 
red-detuned wing is attributed to the nonsharp edges of the electric-field 
switching in the experiment. The resonance amplitude saturates at about 0.125 
value [Fig.3(b)] instead of 0.25, because half of the Rydberg atoms are 
excited to the noninteracting $37P_{1/2} $ state. The dependence of the 
amplitude on \textit{N} shows the good agreement between theory and 
experiment. The experimental dependence of the resonance FWHM on \textit{N} 
is also close to theory [Fig.3(c)]. 

To conclude, we have observed for the first time the F$\ddot{\mathrm{o}}$rster resonance between two Rydberg atoms. 
Although the atom positions were not fixed, the interaction strength, the signal-to-noise ratio, and the spectral 
resolution were large enough for the line-shape analysis.
The line shape agrees well with theory, showing that the two-atom interactions are controlled
by an electric field in a predictable way. This is an important step towards implementation of the electrically controlled 
neutral-atom quantum gates. The next step should be the observation of coherent population oscillations at the
F$\ddot{\mathrm{o}}$rster resonance, which is necessary for quantum phase gates [16].

We appreciate fruitful discussions with E.~Arimondo and M.~Saffman. 
This work was supported by the RFBR (Grants No. 09-02-90427 and No. 09-02-92428) jointly with the Consortium EINSTEIN, 
by the Russian Academy of Sciences, and by the Dynasty Foundation.

\end{document}